\documentclass[manuscript,screen,acmsmall]{acmart}

\usepackage{multirow}
\usepackage{graphicx}
\usepackage{caption}
\usepackage{float}
\usepackage{subfigure}
\usepackage{subcaption}
\usepackage{array}
\usepackage{makecell}
\usepackage[most]{tcolorbox}
\usepackage{booktabs} 
\usepackage{longtable} 
\usepackage{ragged2e} 
\usepackage{needspace} %
\usepackage{enumitem}

\AtBeginDocument{%
  }

\setcopyright{acmlicensed}

\acmISBN{978-1-4503-XXXX-X/24/09}

\begin{document}

\title{Internal Vulnerabilities, External Threats: A Grounded Framework for Enterprise Open Source Risk Governance}


\author{Wenhao Yang}
\orcid{0000-0002-1005-1974}
\affiliation{%
  \institution{Peking University}
  \city{Beijing}
  \country{China}
}
\email{yangwh@stu.pku.edu.cn}

\author{Minghui Zhou}
\orcid{0000-0001-6324-3964}
\authornote{Corresponding author.}
\affiliation{%
  \institution{Peking University}
  \city{Beijing}
  \country{China}
}
\email{zhmh@pku.edu.cn}

\author{Daniel Izquierdo Cortázar}
\affiliation{%
  \institution{Bitergia}
  \city{Madrid}
  \country{Spain}
}
\email{dizquierdo@bitergia.com}

\author{Yehui Wang}
\affiliation{%
  \institution{Huawei Technologies Co., Ltd.}
  \city{Shanghai}
  \country{China}
}
\email{wangyehui2@huawei.com}


\begin{abstract}
Enterprise engagement with open source has evolved from tactical adoption to strategic deep integration, exposing them to a complex risk landscape far beyond mere code. 
However, traditional risk \textit{management}, narrowly focused on technical tools, is structurally inadequate for systemic threats like upstream ``silent fixes'', community conflicts, or sudden license changes, creating a dangerous governance blind spot. 
To address this governance vacuum and enable the necessary shift from tactical risk \textbf{management} to holistic risk \textbf{governance}, we conducted a grounded theory study with 15 practitioners to develop a holistic risk governance framework.

Our study formalizes an analytical framework built on a foundational risk principle: an uncontrollable \textbf{External Threat} (e.g., a sudden license change in a key dependency) only becomes a critical risk when it exploits a controllable \textbf{Internal Vulnerability} (e.g., an undefined risk appetite for single-vendor projects), which then amplifies the impact. 
The framework operationalizes this principle through a clear logical chain: ``\textbf{Objectives $\rightarrow$ Threats $\rightarrow$ Vulnerabilities $\rightarrow$ Mitigation}'' (OTVM). This provides a holistic decision model that transcends mere technical checklists. Based on this logic, our contributions are: 
(1) a ``Strategic Objectives Matrix'' to clarify goals; 
(2) a systematic dual taxonomy of External Threats (\texttt{Ex-Tech}, \texttt{Ex-Comm}, \texttt{Ex-Eco}) and Internal Vulnerabilities (\texttt{In-Strat}, \texttt{In-Ops}, \texttt{In-Tech}); 
and (3) an actionable mitigation framework mapping capability-building to these vulnerabilities. 
The framework's analytical utility was validated by three industry experts through retrospective case studies on real-world incidents. 
This work provides a novel diagnostic lens and a systematic path for enterprises to shift from reactive ``firefighting'' to proactively building an organizational ``immune system''.
\end{abstract}




\maketitle

\section{Introduction}
Open source software has become the bedrock of modern digital infrastructure. As the recent \texttt{XZ Utils} backdoor incident starkly revealed, the nature of open source risk has evolved far beyond code defects to the exploitation of community trust~\cite{cve_xz, orojcik2024xz, bissyande2024xz}.
For enterprises, reliance on and participation in open source have shifted from tactical adoption to strategic deep integration~\cite{nagle2022open}. Today, they are not merely ``consumers''; they are ``contributors'', ``producers'', and ``ecosystem orchestrators'', leveraging open source to gain technology leadership, build business moats, and win the war for talents~\cite{widenius2014business}.
This strategic shift is accompanied by complex risks that traditional risk \textbf{\textit{management}}, often narrowly focused on operational tools like vulnerability scanning, is structurally inadequate for.
This inadequacy creates a dangerous \textbf{\textit{governance}} blind spot, exposing firms to a landscape of profound, yet often unmanaged, pain points.

To diagnose the root of this inadequacy, we explicitly distinguish between \textit{risk management} and \textit{risk governance}. Following established literature in both management theory \cite{weill2004it} and industry practice frameworks like COBIT \cite{isaca2012cobit}, we define management as the operational execution of controls—the processes and tools aimed at ``doing things right''.
Governance, in contrast, is the strategic framework of policies, risk appetite, and accountability set by leadership to ensure the organization is ``doing the right things''. We argue these pain points are not mere failures of management, but symptoms of a deeper \textbf{\textit{governance vacuum}}. This vacuum is the central problem we address, as it leads enterprises to a critical error: they conflate largely uncontrollable \textbf{External Threats} with controllable \textbf{Internal Vulnerabilities}, thereby amplifying impact.

For example, a company may fear its products are exposed due to an upstream ``silent fix''—an urgent security patch without a Common Vulnerabilities and Exposures (CVE) identifier—because it lacks the internal capability for proactive community monitoring, relying solely on lagging public vulnerability databases.
Another firm may struggle to contribute to a upstream  project, only to be told to ``use your own fork'', forcing it into costly maintenance due to an immature ``upstream-first'' culture.
A third might face a strategic crisis when a core dependency's license is unilaterally changed to a restrictive one, a threat amplified by an undefined risk appetite for single-vendor-dominated projects.
Together, these cases span external threats across technology, community, and ecosystem dynamics, and internal vulnerabilities in technology, operations, strategy.

Viewed through the risk lifecycle, what persists across these episodes is not the absence of tools but the misalignment between external triggers and internal amplifiers.
Following modern risk theory \cite{arbaugh2000windows}, we argue that these disparate struggles share a common mechanism: a persistent confusion over the linkage between threats and vulnerabilities. This confusion, we contend, originates from the governance vacuum.
We therefore define \textbf{Internal Vulnerabilities} as controllable organizational weaknesses—beyond technical defects—across strategic, operational, and technical dimensions that external threats can exploit to amplify impact.

This governance vacuum is also reflected in the literature, leading to a critical knowledge gap. While risks are studied at the technical \cite{zhao2023sca, charoenwet2024sast, gao2023characterizing, tan2022exploratory, xu2023understanding, wu2024large, kalu2025industry, ladisa2024sok}, community \cite{zhou2015will, xiao2023how, freund2024backdoor, orojcik2024xz, zhou2017scalability, tan2020scaling, zhou2012investigating, zhou2011initial, zhou2010growth, zhou2010developer}, and ecosystem \cite{ma2013commercial, zhang2022commercial, zhang2021companies, zhang2020how, zhou2016inflow, osborne2024characterising, capra2008open, zaitsev2024opencore} levels, they are often examined in isolation. Consequently, this research remains \textit{fragmented}, lacking a unifying analytical framework.
To bridge this fragmented landscape, we propose an analytical governance framework built upon the foundational relationship between threats and vulnerabilities. We refer to its four-step logic as the ``Objectives $\rightarrow$ Threats $\rightarrow$ Vulnerabilities $\rightarrow$ Mitigation'' chain (OTVM), which we operationalize through the following four fundamental and interconnected questions (RQs):

\textbf{RQ1: What strategic objectives do enterprises pursue through open source activities?}
\textit{Motivation:} Any effective risk governance must be anchored in the value it creates and protects. Without a clear strategic intent, risk analysis becomes an aimless technical exercise. Therefore, the first step of governance must be to clarify the diverse strategic intentions behind an enterprise's engagement in open source.

\textbf{RQ2: What external threats endanger these strategic objectives?}
\textit{Motivation:} The universe of external threats is vast and ever-changing. This research does not aim to create an encyclopedic list of every possible threat. Instead, its objective is to develop a \textbf{taxonomy}—a structured classification system—that organizes this chaos into meaningful categories. This classification serves as the essential first step for the framework's core diagnostic function.

\textbf{RQ3: What internal vulnerabilities within the enterprise amplify the impact of external threats?}
\textit{Motivation:} Having established the role of internal weaknesses as risk amplifiers, this question seeks to systematically classify these organizational ``internal troubles'' as defined above. This diagnosis is the key to finding the root causes of recurring problems and shifting from reactive firefighting to proactive capability building.

\textbf{RQ4: How do enterprises systematically mitigate risks by addressing their vulnerabilities?}
\textit{Motivation:} Ultimately, this research aims to provide an actionable framework that moves beyond a simple checklist of ``best practices''. It guides enterprises in building organizational resilience by systematically mapping mitigation strategies to a clear diagnosis of internal vulnerabilities.

To answer these questions, we employ a rigorous Grounded Theory methodology~\cite{glaser1967discovery}, developing a new framework for enterprise open source risk governance directly from interviews with 15 industry practitioners. This novel, empirically grounded framework's primary contributions include:
\begin{itemize}
    \item The conceptualization and formalization of the ``Objectives $\rightarrow$ Threats $\rightarrow$ Vulnerabilities $\rightarrow$ Mitigation'' logical chain (OTVM) as a new analytical lens for the field.
    \item A Strategic Objectives Matrix that systematically maps the diverse value-creation pathways of enterprise open source engagement.
    \item An empirically grounded, three-pillar External Threat Taxonomy (\texttt{Ex-Tech}, \texttt{Ex-Comm}, \texttt{Ex-Eco}) and a corresponding Internal Vulnerability Taxonomy (\texttt{In-Strat}, \texttt{In-Ops}, \texttt{In-Tech}).
    \item An \textit{actionable} Risk Mitigation Framework that links mitigation strategies to diagnosed internal vulnerabilities, forming a capability-building roadmap.
    \item Validation of the framework's analytical utility through retrospective case studies conducted with three industry experts.
\end{itemize}

This framework equips enterprises with a novel \textit{diagnostic lens} and an actionable guide for building a robust governance system. It shifts the focus from trying to control an uncontrollable external world to systematically building an internal ``immune system''.

\section{Related Work}
The challenge of managing open source software risks is not new. However, the existing body of research, while extensive, is fragmented. This fragmentation is not merely a characteristic of the literature but a direct reflection of the central problem this paper addresses: the lack of a unifying analytical framework. By examining technical, community, and ecosystem risks in isolated silos, current research mirrors and reinforces the very same fragmented governance that leads to misaligned, reactive decision-making in practice.

\subsection{Multi-dimensional Research on Open Source Risks}
The literature on open source risks, while extensive, has recently matured to a point where systematic overviews are emerging. Comprehensive surveys now map the landscape of attacks, risk assessment strategies, and security controls \cite{Gokkaya2023Software}, while influential Systematization of Knowledge (SoK) studies provide detailed, hierarchical taxonomies of specific attack vectors targeting the software supply chain \cite{Ladisa2023SoK}. These foundational works confirm the multi-faceted nature of the challenge, which we analyze across three interconnected streams.

\paragraph{Technical \& Product Risks} This is the most mature research area, focusing on the software artifacts themselves. Foundational work on Software Composition Analysis (SCA) aims to detect vulnerabilities and license compliance issues. However, recent large-scale empirical studies confirm significant accuracy challenges with existing tools. For instance, a 2023 study on Java projects found the average F1 score for vulnerability detection to be as low as 0.475 \cite{zhao2023sca}, while another 2024 study on C/C++ found that over 76\% of tool warnings in vulnerable functions were unrelated to the actual vulnerability-introducing change \cite{charoenwet2024sast}. Recent studies have expanded this view by characterizing the complex supply chains of specific domains, such as Deep Learning packages in the Python Package Index (\texttt{PyPI}), revealing their unique structures and risks \cite{gao2023characterizing, tan2022exploratory}. A significant technical risk lies in license compliance. Researchers have deeply investigated license incompatibilities, such as the prevalence and remediation challenges of incompatible licenses in large ecosystems like \texttt{PyPI} \cite{xu2023understanding}, highlighting the legal complexities enterprises face \cite{wu2024large}. To improve supply chain transparency and combat these risks, generating an accurate Software Bill of Materials (SBOM) is critical, but its practical implementation faces deep socio-technical barriers \cite{kalu2025industry}. A recent comprehensive Systematization of Knowledge (SoK) on SBOMs confirms this, identifying a wide array of non-technical barriers to adoption, including issues with generation tools, data privacy, maintenance overhead, and a lack of effective analysis tools \cite{ladisa2024sok}.

\paragraph{Socio-technical \& Community Risks} A growing body of research focuses on the ``people'' behind the code, recognizing that community health is a direct precursor to technical quality and security. Studies have modeled what makes contributors stay in a community long-term~\cite{zhou2015will}, and how early participation patterns can predict sustained activity \cite{xiao2023how}.
This focus on human factors \cite{zhou2012investigating} extends to the entire developer lifecycle, from the impact of their initial environment \cite{zhou2011initial} and the growth of newcomer competence \cite{zhou2010growth} to their overall developer fluency \cite{zhou2010developer}. 
The way communities are structured and scaled also presents challenges and requires specific practices for decentralization to succeed~\cite{tan2022scaling}, as observed in large-scale projects like the Linux kernel, where both community \cite{tan2020scaling} and maintainer scalability \cite{zhou2017scalability} are critical challenges. 
Understanding these human factors is vital, as maintainer burnout and lack of resources can lead to project stagnation and create security risks \cite{linuxfoundation2022maintainer}. The \texttt{XZ Utils} backdoor \cite{cve_xz} was a watershed moment, demonstrating that a technical vulnerability can be the end result of a multi-year social engineering attack that exploits community trust and maintainer fatigue~\cite{freund2024backdoor, orojcik2024xz}. 

\paragraph{Ecosystem \& Macro-Environment Risks} This stream extends the view to the broader landscape where enterprises and communities interact, often with complex dynamics of cooperation and competition (``co-opetition''). The significant role of commercial participation in large-scale open source ecosystems like \texttt{OpenStack} has been extensively studied, revealing it to be a double-edged sword that can both vitalize and dominate a project \cite{zhang2022commercial, zhang2021companies}, profoundly shaping how companies collaborate \cite{zhang2020how}, impacting contributor inflow and retention \cite{zhou2016inflow}, and even influencing community processes like issue reporting \cite{ma2013commercial}. A recent study on company-hosted projects like \texttt{PyTorch} and \texttt{TensorFlow} further characterizes this co-opetition, analyzing how corporations interact within ecosystems they nominally control~\cite{osborne2024characterising}. These ecosystem dynamics can lead to risks such as forking due to strategic license changes, often driven by the ``Open Core'' business model \cite{capra2008open, zaitsev2024opencore}. More recently, macro-level risks from global regulatory pressures, such as the EU's Cyber Resilience Act (CRA), have introduced new legal compliance challenges for all software producers \cite{redhat2024cra}.

\subsection{Research Gap and Our Contribution}
As our literature review demonstrates, the academic understanding of open source risk is itself fragmented and imbalanced. Research into \textbf{Technical \& Product Risks} is by far the most mature, with influential taxonomies now providing systematic classification of the external attack surface \cite{Gokkaya2023Software, Ladisa2023SoK}. This academic focus on external threats directly corresponds to the prevalent industry practice of ``traditional governance'', which is narrowly centered on vulnerability scanning and license compliance. However, the growing bodies of work on \textbf{Socio-technical} and \textbf{Ecosystem} risks reveal the profound inadequacy of this approach.

This imbalance sharpens the critical research gap: our ability to catalog external threats has far outpaced our ability to systematically analyze the \textit{internal vulnerabilities} that determine their ultimate impact. Incidents like the \texttt{XZ} backdoor are not merely external events; their destructive power is unlocked by internal frailties in community trust and maintainer well-being that traditional SCA tools cannot detect. Similarly, ecosystem-level shifts like strategic license changes pose business continuity risks far greater than a single CVE. This disconnect between the holistic nature of the risk and the narrow, tool-focused analysis leads directly to the \textbf{misaligned, reactive decision-making} that plagues many enterprises.

Our primary contribution is to correct this imbalance and bridge the resulting gap. While prior work maps parts of the external landscape, our work synthesizes these known challenges into a coherent, \textbf{analytical governance model} that provides the missing internal perspective. Through a rigorous grounded theory process, we formalize the ``Objectives → Threats → Vulnerabilities → Mitigation'' logical chain. This model provides a novel and actionable lens for enterprises to move from reactively fighting external fires to proactively building an internal ``immune system''.

\section{Methodology}
To construct a governance framework deeply rooted in industrial practice, this study employs a qualitative research methodology based on Grounded Theory \cite{glaser1967discovery}, following best practices for its application within software engineering research \cite{stol2016grounded}. The research is divided into two core phases: 
\textbf{(1) Framework Construction}: Using data from 15 practitioners (P1-P15) to generate concepts and build the framework from the ground up via open, axial, and selective coding. 
\textbf{(2) Framework Validation}: Using 3 independent experts (E1-E3) to evaluate the final integrated framework's \textit{analytical utility} through retrospective case studies.
It is crucial to distinguish our \textit{methodology} from our \textit{contribution}: we use Grounded Theory (the methodology) to build an empirically-grounded governance framework (the contribution), not a predictive ``theory'' in the classical sense.

\subsection{Framework Construction}
We followed the guiding principles of Grounded Theory~\cite{glaser1967discovery}, a bottom-up research strategy aimed at generating theory directly from raw data, rather than verifying preconceived hypotheses. This approach is particularly suitable for exploring complex domains like enterprise open source risk where comprehensive, empirically validated theories are lacking.

\subsubsection{Data Collection and Analysis}
The core of our framework construction was an iterative process where data collection and analysis were interwoven, consistent with Grounded Theory principles.
The 15 practitioners in this phase (P1-P15) served as the primary data source for \textit{inductive concept generation}.

\paragraph{Theoretical Sampling and Participant Selection} 
We employed Theoretical Sampling, a strategy where data collection is guided by the emerging theory itself. The goal is not to maximize diversity for its own sake, but to purposefully select participants and data that can help elaborate on, refine, and saturate the developing theoretical categories. We interviewed 15 practitioners from 9 different technology companies. This sampling process was iterative: the first few interviewees helped us establish initial concepts, while subsequent selections were intentionally targeted at specific roles and contexts that could challenge, extend, or refine our emerging theoretical constructs.

These participants hold diverse roles within their companies, including Open Source Program Office (OSPO) leaders \cite{todo_ospo_definition}, software supply chain security experts, head of Research and Development (R\&D), community operation managers, AI software stack architects, and core engineers.

\begin{table}[h]
\centering
\caption{Interview Participants for Framework Construction}
\label{tab:participants}
\small
\renewcommand{\arraystretch}{1.2}
\begin{tabular}{l l l}
\toprule
\textbf{ID} & \textbf{Position} & \textbf{Company Domain (and name)} \\
\midrule
P1 & Principal Engineer, AI Frameworks & AI Chip Manufacturer (A) \\
P2 & Software Architect, AI Stack & AI Chip Manufacturer (A) \\
P3 & Software Architect, AI Stack & AI Chip Startup (B) \\ \hline
P4 & Director, Open Source Strategy & \makecell[l]{Cloud Service Provider (C)} \\
P5 & Manager, OSPO Governance & \makecell[l]{Cloud Service Provider (C)} \\
P6 & Lead, Open Source Standards & \makecell[l]{Cloud Service Provider (C)} \\
P7 & Senior Cloud Native Engineer & Cloud Service Provider (D) \\ \hline
P8 & Senior Data Analyst, OSPO & FinTech Company (E) \\ \hline
P9 & Senior Manager, Community Operations & \makecell[l]{Internet Service Provider (F)} \\
P10 & Director, Open Source Ecosystem & \makecell[l]{Internet Service Provider (F)} \\ \hline
P11 & Principal Security Engineer & Operating System Vendor (G) \\
P12 & Senior Software Engineer, Framework Integration & Operating System Vendor (G) \\
P13 & Head of Infrastructure \& Operations & Operating System Vendor (G) \\ \hline
P14 & AI Application Engineer & \makecell[l]{Traditional Industry (Transportation) (H)} \\
P15 & Head of R\&D & \makecell[l]{Traditional Industry (Construction) (I)} \\
\bottomrule
\end{tabular}
\end{table}

\paragraph{Semi-structured Interviews} The primary method of data collection was semi-structured interviews, each lasting 60 to 90 minutes. We designed an interview guide covering core topics such as the interviewee's background, open source engagement scenarios, challenges encountered, decision-making processes, and risk perception. The semi-structured format allowed us to follow the guide to ensure core questions were covered while flexibly asking follow-up questions based on the interviewee's responses to delve deeper into unexpected insights and details. All interviews were audio-recorded with the participants' consent and transcribed for subsequent analysis.

\paragraph{Data Analysis: Emergence of the Core Theory}
We rigorously followed the three-stage coding process of Grounded Theory to analyze the interview transcripts, continuously using the Constant Comparison method to test and refine our theory. This systematic process is precisely how we uncovered the intrinsic connections between the research questions posed in our introduction.

\begin{enumerate}
    \item \textbf{Open Coding:} In the first stage, two researchers independently analyzed the text line-by-line, breaking it down into discrete concepts and assigning initial labels (e.g., ``fear of upstream abandonment'', ``slow internal legal review''). Regular meetings were held to discuss and align these codes, ensuring inter-coder reliability.

    \item \textbf{Axial Coding: Emergence of the ``Internal/External'' Dualism.} As we categorized the concepts from open coding, a powerful and persistent pattern emerged from the data. We found that practitioners' descriptions of their challenges naturally clustered into two major groups. The first group of codes, such as ``upstream PR rejected'', ``maintainer disappeared'', or ``competitor changed license'', shared a common characteristic: the problem's source was in the external open source ecosystem, beyond the enterprise's direct control. The second group, including codes like ``internal approval process too slow'', ``engineers won't write docs'', or ``security and dev teams conflict'', shared a different characteristic: the problem's source was the enterprise's own organizational, process, or cultural deficiencies. This clear ``internal/external attribution'' dichotomy, emerging repeatedly from the data itself rather than from any preconceived model, became a fundamental pillar of our theory. It was this emergent dualism that provided the first core insight into how enterprises structure their risk perception, establishing a conceptual link between our research questions on external threats (RQ2) and internal vulnerabilities (RQ3).

    \item \textbf{Selective Coding: Emergence of the Four-Step Governance Logic.} In the final stage, our goal was to integrate all core categories into a coherent analytical storyline. We traced the causal relationships in practitioners' narratives of incidents and governance activities. Ultimately, a complete governance storyline repeatedly emerged, one that begins with objectives, proceeds through external threats, is attributed to internal vulnerabilities, and concludes with mitigation. At this point, the four seemingly separate research questions posed in our introduction were integrated by the data itself into an indivisible theoretical framework with an inherent logical order: ``Objectives → Threats → Vulnerabilities → Mitigation''. This logical chain was not something we presumed; it is the core conceptual discovery that emerged from our research.
\end{enumerate}

\subsection{Framework Validation}
\textbf{Crucially, their role was distinct from the P1-P15 participants: their purpose was not to serve as a data source for coding, but to \textit{apply} the fully constructed framework as domain experts.}To evaluate our emergent framework, we conducted independent external validation with 3 industry experts (labeled E1-E3) who had not participated in the initial interviews. To move beyond a simple assessment of face validity (``does this look reasonable?''), our validation process included a key activity: we asked each expert to select a significant, real-world open source-related event from their recent experience and use our framework's concepts to conduct a \textbf{retrospective case study}. Our goal was to assess the framework's \textit{analytical utility}—its ability to structure complex events and reveal deeper insights that might have been missed at the time.

\begin{table}[h]
\centering
\caption{Expert Participants for Framework Validation}
\label{tab:validators}
\small
\renewcommand{\arraystretch}{1.2}
\begin{tabular}{l l l}
\toprule
\textbf{ID} & \textbf{Position} & \textbf{Company Domain (and name)} \\
\midrule
E1 & Director, OSPO & FinTech Company (J) \\ \hline
E2 & Principal Security Architect & Cloud Service Provider (K) \\ \hline
E3 & Head of R\&D & AI Chip Startup (L) \\
\bottomrule
\end{tabular}
\end{table}

\section{An Grounded Framework for Enterprise OSS Risk Governance}
Our grounded theory analysis yielded a comprehensive framework for enterprise open source risk governance. This framework is composed of four logically connected parts, corresponding to our research questions.

\subsection{RQ1: The Enterprise Open Source Strategy Matrix} 
The logical starting point of our theory is the enterprise's strategic objectives. We organized these into a two-dimensional matrix as shown in Table~\ref{tab:strategy-matrix}. The horizontal axis outlines five common ``Open Source Engagement Scenarios'' an enterprise might adopt. 
The vertical axis represents five ``Value Dimensions'', the core areas where benefits are expected. Each cell in the matrix is coded (e.g., [1.1]) to correspond with its detailed description in Section 4.1.3.

\subsubsection{Open Source Engagement Scenarios (X-axis)}
We define five core engagement scenarios that enterprises adopt. Crucially, these are not a strict, linear maturity ladder, but rather a portfolio of strategies that a large enterprise may employ concurrently depending on business needs. 
\begin{enumerate}
    \item \textbf{Consume:} The foundational scenario. The enterprise uses external open source software to accelerate development and reduce R\&D costs.
    \item \textbf{InnerSource:} The enterprise applies open source principles (e.g., open collaboration, code reuse) to its internal software development to break down silos and improve engineering efficiency. It is a key bridge from consumption to contribution.
    \item \textbf{Contribute:} The enterprise contributes resources to strategically important external projects to influence their direction, reduce the maintenance cost of internal forks, and build its technical reputation.
    \item \textbf{Produce:} The enterprise initiates and leads its own open source projects, typically by open-sourcing an internal tool, to build a developer ecosystem around its technology.
    \item \textbf{Orchestrate:} The highest level of engagement. The enterprise leads a broad, multi-party technology ecosystem, often through a neutral foundation, to define and steer industry-level platform standards.
\end{enumerate}

\subsubsection{Value Dimensions (Y-axis)}
\begin{itemize}
    \item \textbf{Dimension 1: Engineering \& Product Excellence:} The direct impact on R\&D efficiency and product quality.
    \item \textbf{Dimension 2: Ecological Niche \& Strategic Advantage:} How open source helps an enterprise build a defensible, long-term competitive moat.
    \item \textbf{Dimension 3: Talent \& Developer Ecosystem:} The impact on attracting and retaining internal talent and building an external developer community.
    \item \textbf{Dimension 4: Business Model \& Value Capture:} The specific paths to convert open source activities into measurable business value, such as revenue or cost savings.
    \item \textbf{Dimension 5: Governance, Security \& Compliance:} The foundational capabilities required to enable and safeguard all other strategic objectives.
\end{itemize}

\begin{table}[t]
\centering
\caption{The Enterprise Open Source Strategy Matrix}
\label{tab:strategy-matrix}
\small
\renewcommand{\arraystretch}{1.5}
\begin{tabular}{
>{\bfseries\RaggedRight}p{0.14\textwidth} 
>{\RaggedRight}p{0.14\textwidth} 
>{\RaggedRight}p{0.145\textwidth} 
>{\RaggedRight}p{0.14\textwidth} 
>{\RaggedRight}p{0.15\textwidth} 
>{\RaggedRight\arraybackslash}p{0.15\textwidth}
}
\toprule
\textbf{Value Dimensions} & \multicolumn{5}{c}{\textbf{Open Source Engagement Scenarios}} \\
\cmidrule{2-6}
& \textbf{Consume} & \textbf{InnerSource} & \textbf{Contribute} & \textbf{Produce} & \textbf{Orchestrate} \\
\midrule
\textbf{Engineering \& Product Excellence} & [1.1] Accelerate R\&D, Reduce TCO & [1.2] Foster Internal Reuse & [1.3] Influence Upstream, Reduce Maint. Cost & [1.4] Standardize Internal Tech Stack & [1.5] Drive Industry-level Tech Breakthroughs \\
\textbf{Ecological Niche \& Strategic Advantage} & [2.1] Avoid Vendor Lock-in & [2.2] Cultivate Internal Innovation & [2.3] Gain Industry Voice & [2.4] Establish De Facto Tech Standards & [2.5] Steer Ecosystem Evolution \\
\textbf{Talent \& Developer Ecosystem} & [3.1] Build Modern Tech Brand & [3.2] Accelerate Talent Growth & [3.3] Establish Global Tech Reputation & [3.4] Build Own Developer Community & [3.5] Become Talent Hub for a Domain \\
\textbf{Business Model \& Value Capture} & [4.1] Reduce Direct Costs & [4.2] Optimize R\&D Resources & [4.3] Support Professional Services Model & [4.4] Enable Diversified Business Models & [4.5] Drive Core Business Growth \\
\textbf{Governance, Security \& Compliance} & [5.1] Manage Supply Chain Risks & [5.2] Promote Internal Security Best Practices & [5.3] Protect Corporate IP & [5.4] Establish Trusted Project Governance & [5.5] Set Ecosystem Security Standards \\
\bottomrule
\end{tabular}
\end{table}

\subsubsection{Matrix Details}
\paragraph{Dimension 1: Engineering \& Product Excellence}
\begin{itemize}[label={}, leftmargin=2em]
    \item\textbf{Consume [1.1] Accelerate R\&D, Reduce TCO:} Utilize existing external components to improve product quality and stability, and reduce the Total Cost of Ownership (TCO).
    A Head of R\&D from a traditional industry (P15) stated:
    \begin{quote}
    \textit{Our team heavily uses code from the open source community. For a new application, we usually spend about two weeks on technical evaluation, including which open source modules and databases to use. This greatly speeds up our development pace.}
    \end{quote}
    \item\textbf{InnerSource [1.2] Foster Internal Reuse:} Break down team silos, promote code reuse, and unify internal best practices.
    \item\textbf{Contribute [1.3] Influence Upstream, Reduce Maintenance Costs:} Merge customized features and fixes into the main branch, eliminating the burden of maintaining internal forks.
    An OSPO Standards Lead (P6) explained:
    \begin{quote}
    \textit{If you benefit greatly from open source, you should give back. Contributing our patches back upstream unloads a heavy burden. In the long run, this will definitely reduce our maintenance costs.}
    \end{quote}
    \item\textbf{Produce [1.4] Standardize Internal Tech Stack:} Turn a self-developed core framework into an internal standard to reduce fragmentation and maintenance overhead.
    An Ecosystem Lead (P10) recalled:
    \begin{quote}
    \textit{In the years after our AI framework was open-sourced, one of our main tasks was to promote unification across the company's internal businesses... Our principle was `no reinventing the wheel within the company'; everyone had to use this framework. The reform process was painful, but the results were good.}
    \end{quote}
    \item\textbf{Orchestrate [1.5] Drive Industry-level Technical Breakthroughs:} Leverage the R\&D power of the global community to solve complex technical problems that a single enterprise cannot tackle alone.
    An OSPO Strategist (P4) believes:
     \begin{quote}
    \textit{Fundamental, platform-level technical challenges like building a next-generation operating system require massive R\&D investment, far beyond the capacity of any single company. By orchestrating a multi-party open source ecosystem, we can gather the world's top wisdom and R\&D resources to solve these problems together.}
    \end{quote}
\end{itemize}

\paragraph{Dimension 2: Ecological Niche \& Strategic Advantage}
\begin{itemize}[label={}, leftmargin=2em]
    \item\textbf{Consume [2.1] Avoid Vendor Lock-in:} Adopt open standards and ecosystems to maintain strategic flexibility and enhance supply chain resilience.
    An engineer from an AI chip manufacturer (P1) pointed out:
    \begin{quote}
    \textit{Why would customers consider us? The reason is simple: you're in a tough spot if you only have one supplier. So, our customers are always looking for a second source, at least to use our pricing to negotiate with the market leader.}
    \end{quote}
    \item\textbf{InnerSource [2.2] Cultivate Internal Innovation:} Create a bottom-up innovation environment, accelerating the prototyping of new products and technologies.
    \item\textbf{Contribute [2.3] Gain Industry Voice:} Become a core member of key projects and foundations to participate in shaping industry standards and technical direction.
    A senior OSPO strategist (P4) shared:
    \begin{quote}
    \textit{How did we get our engineers to become core members of a key container orchestration project? It was all orchestrated. We sent people to the cities where core developers gathered for long periods, specifically to build relationships and sponsor their meetups. That's how we gradually gained a voice in the community.}
    \end{quote}
    \item\textbf{Produce [2.4] Establish De Facto Tech Standards:} Promote a self-developed project to become an industry standard, creating a technological moat and defining a new market category.
    An engineer at an AI chip startup (P3) noted:
     \begin{quote}
    \textit{The market leader has been operating for twenty years and has built a powerful ecosystem moat. The biggest pain point for us latecomers is that we have to follow their standards for everything.}
    \end{quote}
    \item\textbf{Orchestrate [2.5] Steer Ecosystem Evolution:} Control the strategic direction of platform-level technologies, commoditizing complements through standardized interfaces to solidify core business advantages.
    A senior engineer (P7) commented:
     \begin{quote}
    \textit{Our leadership in the open source cloud-native stack isn't about directly selling that stack. Our goal is to standardize and commoditize the application development and deployment layer. By doing so, the unique value of our underlying cloud platform—like better performance and lower cost—becomes much more apparent.}
    \end{quote}
\end{itemize}

\paragraph{Dimension 3: Talent \& Developer Ecosystem}
\begin{itemize}[label={}, leftmargin=2em]
    \item\textbf{Consume [3.1] Build Modern Tech Brand:} Adopting mainstream open source technology stacks increases attractiveness to top engineering talent.
    A senior engineer (P7) stated:
    \begin{quote}
    \textit{Top engineers don't want to work with outdated proprietary technology. We've found in recruiting that simply stating our stack is based on \texttt{Kubernetes}, \texttt{Go}, and \texttt{Rust} is a powerful signal of attraction.}
    \end{quote}
    \item\textbf{InnerSource [3.2] Accelerate Talent Growth:} Provide cross-project collaboration opportunities, promote internal knowledge flow, and enhance engineers' comprehensive abilities.
    \item\textbf{Contribute [3.3] Establish Global Tech Reputation \& Recruit Talent:} Showcase technical prowess in top-tier projects to precisely identify and recruit exceptional contributors from the community.
    A community operations manager (P9) mentioned:
    \begin{quote}
    \textit{Among the core community members I track, about 10\% have become full-time employees or interns at our company. Their community contributions provide another dimension for evaluation during interviews, significantly increasing our hiring success rate.}
    \end{quote}
    \item\textbf{Produce [3.4] Build Own Developer Community:} Build a talent pool around a self-developed project to get direct user feedback and external innovation.
    The same community manager (P9) shared:
     \begin{quote}
    \textit{Through a series of activities, we get developers involved in our daily development. Internal R\&D teams that have experienced the benefits of community contributions are very willing to publish development tasks, because it makes development incredibly fast.}
    \end{quote}
    \item\textbf{Orchestrate [3.5] Become Talent Hub for a Domain:} Become the ``center of gravity'' for a specific technology domain, attracting the world's top talent to your ecosystem.
    An OSPO Strategist (P4) asserted:
     \begin{quote}
    \textit{When you orchestrate an ecosystem like \texttt{Kubernetes}, you're not just defining the technology; you're defining the job market for it. The best cloud-native talent worldwide is drawn into your ecosystem, giving you an unparalleled recruiting advantage.}
    \end{quote}
\end{itemize}

\paragraph{Dimension 4: Business Model \& Value Capture}
\begin{itemize}[label={}, leftmargin=2em]
    \item\textbf{Consume [4.1] Reduce Direct Costs \& Entry Barriers:} Save on software licensing fees and meet procurement requirements in specific markets (e.g., for code auditability).
    A Head of R\&D from a traditional industry (P15) mentioned:
    \begin{quote}
    \textit{For some of the government projects we work on, there is a `secure and controllable' consideration, which requires us to prioritize using domestic open source products.}
    \end{quote}
    \item\textbf{InnerSource [4.2] Optimize R\&D Resources:} Reduce resource waste from internal "reinventing the wheel," focusing engineering efforts on core business value.
    \item\textbf{Contribute [4.3] Support Professional Services Model:} Leverage deep upstream contributions to build an expert reputation, driving high-value enterprise support and consulting services.
    A senior engineer (P7) explained:
    \begin{quote}
    \textit{Our deep contributions to the upstream project are not charity. They build our expert credibility. When a major financial institution has a problem with a critical system, they don't need a regular consultant; they need us, the people who actually write and maintain the core code. That's the source of our high-value enterprise support services.}
    \end{quote}
    \item\textbf{Produce [4.4] Enable Diversified Business Models:} Build various monetization paths around a self-developed project, such as open-core, SaaS, or dual-licensing.
    An OSPO Strategist (P4) elaborated on the logic of the ``Open Core'' model \cite{capra2008open, zaitsev2024opencore}:
     \begin{quote}
    \textit{We open-source a feature-complete community edition to maximize user adoption and build an ecosystem. When users deploy it at scale in production and develop strong needs for stability, security, and advanced management features, they naturally become paying customers for our enterprise edition. Open source is the best tool for market education and user qualification.}
    \end{quote}
    \item\textbf{Orchestrate [4.5] Drive Core Business Growth:} Dramatically expand the potential market for core commercial products (like cloud computing or hardware) by leading a free, open source platform.
    Engineers from all the AI chip companies (P1, P2, P3) clearly stated:
     \begin{quote}
    \textit{that the fundamental purpose of leading an open source AI software stack is to drive market adoption and sales of their core commercial product (Graphics Processing Unit (GPU) hardware).}
    \end{quote}
\end{itemize}

\paragraph{Dimension 5: Governance, Security, \& Compliance}
\begin{itemize}[label={}, leftmargin=2em]
    \item\textbf{Consume [5.1] Manage Software Supply Chain Risks:} Establish processes for license scanning, vulnerability detection (SCA), and SBOM management.
    An OSPO Governance Lead (P5) emphasized:
    \begin{quote}
    \textit{When we use open source internally, we have certain management requirements, including component selection and evaluation, as well as subsequent lifecycle maintenance.} The goal is to establish a comprehensive process covering these areas.
    \end{quote}
    \item\textbf{InnerSource [5.2] Promote Internal Security Best Practices:} Improve the overall security of internal code by sharing and reusing pre-vetted security components and libraries.
    \item\textbf{Contribute [5.3] Protect Corporate IP:} Implement clear contribution policies, developer training, and management of contributor agreements like the Contributor License Agreement (CLA) or Developer Certificate of Origin (DCO).
    An AI framework engineer (P1) mentioned:
    \begin{quote}
    \textit{When we participated in an external collaboration project early on, our legal department required every developer to sign a CLA to protect the intellectual property of both companies.}
    \end{quote}
    \item\textbf{Produce [5.4] Establish Trusted Project Governance:} Choose appropriate open source licenses and establish a transparent governance model and a secure vulnerability disclosure process.
    An Ecosystem Lead (P10) stated:
     \begin{quote}
    \textit{Our project's website has a clear Security Policy, and we also coordinate with the CVE process. We follow a very formal security disclosure and response model.}
    \end{quote}
    \item\textbf{Orchestrate [5.5] Set Ecosystem Security Standards:} Lead and coordinate security standards and major vulnerability incident response mechanisms for the entire ecosystem.
    A supply chain security expert (P11) described their role:
     \begin{quote}
    \textit{My scope of responsibility is very broad; it's all `supply chain'. So, in theory, we participate in and communicate about community open source operations, security standards, and so on.} This reflects that the company's goal is to maintain the health and security of the entire ecosystem.
    \end{quote}
\end{itemize}

\subsubsection{Interpreting the Matrix}
The relationship between the axes is one of ``enablement and facilitation'', not a rigid correspondence. Analyzing a column reveals the ``toolbox'' of a single scenario, while analyzing a row reveals the ``maturity roadmap'' for a single goal. The entire matrix serves as a ``strategic decision map'' for diagnosis and planning.

\subsection{RQ2: The External Threat Taxonomy}
The second component of our theory is the classification of external threats. These emerged from the data as three interconnected layers, progressing from direct technical issues (\texttt{Ex-Tech}) to social root causes (\texttt{Ex-Comm}) and finally to the broadest environmental drivers (\texttt{Ex-Eco}). They often form a logical chain.

\begin{table*}[!t]
\centering
\caption{Taxonomy of External Threats}
\label{tab:external-risks}
\small
\begin{tabular}{
    >{\RaggedRight}p{0.10\textwidth} 
    >{\RaggedRight}p{0.25\textwidth} 
    >{\RaggedRight\arraybackslash}p{0.57\textwidth}
}
\toprule
\textbf{ID} & \textbf{Threat Name} & \textbf{Description} \\
\midrule
\multicolumn{3}{c}{\textbf{Technical Threat (\texttt{Ex-Tech})}} \\
\midrule
\texttt{Ex-Tech-1} & Vulnerabilities in Third-Party Dependencies & Technical flaws in components that can be exploited, including known (CVEs) and unknown vulnerabilities, often in transitive dependencies. \\
\texttt{Ex-Tech-2} & Malicious Packages and Commits & Harmful code intentionally introduced by attackers, either via malicious packages (dependency confusion, typosquatting) or through direct commits to a project. \\
\texttt{Ex-Tech-3} & Upstream Breaking Changes & The release of new versions with backward incompatible changes, which can lead to high adaptation costs for enterprises lacking robust testing. \\
\texttt{Ex-Tech-4} & Incompatible or Reciprocal Licenses & The legal risks arising from license terms, particularly strong copyleft (reciprocal) licenses that may compel the open-sourcing of proprietary code. \\
\midrule

\multicolumn{3}{c}{\textbf{Community Threat (\texttt{Ex-Comm})}} \\
\midrule
\texttt{Ex-Comm-1} & Project Abandonment or Stagnation & A depended-upon project is no longer actively maintained (abandoned) or updated (stagnating), creating security and maintenance burdens. \\
\texttt{Ex-Comm-2} & Maintainer Burnout & Critical project knowledge is concentrated in a few maintainers, whose departure due to burnout could cause the project to stall. \\
\texttt{Ex-Comm-3} & Collaboration Barriers & The enterprise encounters indifference or rejection when contributing to upstream communities, forcing the maintenance of costly internal forks. \\
\midrule

\multicolumn{3}{c}{\textbf{Ecosystem Threat (\texttt{Ex-Eco})}} \\
\midrule
\texttt{Ex-Eco-1} & Regulatory Pressure & Governments impose new, legally binding requirements for software supply chain security (e.g., EU's CRA). \\
\texttt{Ex-Eco-2} & Competitive Disadvantage & The risk of market marginalization for a newcomer in a field already dominated by a \textit{de facto} standard (e.g., \texttt{CUDA}). \\
\texttt{Ex-Eco-3} & Ecosystem Fragmentation & A project is forked due to commercial competition or protocol changes, leading to a split in the community and ecosystem. \\
\texttt{Ex-Eco-4} & Strategic License Changes & The commercial entity behind a project unilaterally changes its license to terms with commercial restrictions (e.g., \texttt{SSPL}). \\
\bottomrule
\end{tabular}
\end{table*}

\subsubsection{Dimension 1: Technical Threats (\texttt{Ex-Tech})}

\paragraph{\texttt{Ex-Tech-1}: Vulnerabilities in Third-Party Dependencies}
Open source components contain technical flaws, either publicly disclosed (CVEs) or yet undiscovered. An OS security expert (P11) highlighted the risk of ``silent fixes'':
\begin{quote}
\textit{We pay close attention to `implicit fixes'. Some upstream communities fix vulnerabilities but don't disclose them as CVEs, quietly patching them in a later version. If you don't upgrade in time, your product will have an unperceived risk exposure for a long time.}
\end{quote}

\paragraph{\texttt{Ex-Tech-2}: Malicious Packages and Commits} Attackers compromise the software supply chain by introducing malicious code, either by publishing malicious packages (e.g., through domain squatting or dependency confusion) or by injecting backdoors into legitimate open source projects via malicious commits. A community manager (P9) shared:
\begin{quote}
\textit{We recently encountered a malicious attack. The attacker first submitted a normal \texttt{PR} fixing a typo... Once they became a contributor and gained the trust of our Continuous Integration (CI) system, their second \texttt{PR} started embedding malicious code during runtime, attempting to attack our system.}
\end{quote}

\paragraph{\texttt{Ex-Tech-3}: Upstream Breaking Changes} Upstream projects release updates with backward incompatible changes. For enterprises lacking robust testing and upgrade strategies, this can translate into high adaptation costs and system failures. A Head of R\&D (P15) said:
\begin{quote}
\textit{A major headache with open source now is its lack of backward compatibility... We had two online security incidents this year, both because a base OS distribution had a cross-version update whose inherent incompatibility caused system files to be lost, bringing down the entire server.}
\end{quote}

\paragraph{\texttt{Ex-Tech-4}: Incompatible or Reciprocal Licenses}
The use of software with certain license terms creates legal risks. This is particularly true for strong copyleft (reciprocal) licenses, such as the General Public License (\texttt{GPL}), which may legally compel an enterprise that integrates such code to open-source its own proprietary products. An OSPO Governance Lead (P5) emphasized:
\begin{quote}
\textit{We have strict restrictions on \texttt{GPLv3}... This isn't a technical issue; it's a business survival issue.}
\end{quote}

\subsubsection{Dimension 2: Community Threats (\texttt{Ex-Comm})}
\paragraph{\texttt{Ex-Comm-1}: Project Abandonment or Stagnation} A depended-on project is abandoned or becomes inactive (stagnating), a phenomenon colloquially known as becoming ``zombie code''. The enterprise must then invest far more internal resources than anticipated to maintain an external codebase. An AI engineer (P14) shared:
\begin{quote}
\textit{We used to rely on a deep learning annotation tool, but after the community developer graduated... no one maintained it. Later, only a specific version of the dependency list saved on my personal computer could successfully install it.}
\end{quote}

\paragraph{\texttt{Ex-Comm-2}: Maintainer Burnout} Core knowledge and permissions are concentrated in a few key developers. Their departure due to immense ``maintainer burnout'' pressure can stall the project, creating a maintenance vacuum that can be exploited by malicious actors. This pressure, often termed ``maintainer burnout'', has been empirically studied as a key factor in project health, driven by unhealthy interactions \cite{Raman2020Stress} and an overwhelming demand for support that outstrips a project's maintenance capacity \cite{Valiev2020External}. This risk of knowledge concentration can be quantitatively assessed using the ``Bus Factor''---the minimum number of individuals that would need to leave a project to cripple it---for which more accurate estimation heuristics are now being developed \cite{piccolo2025busfactor}. Furthermore, the onset of burnout is often preceded by detectable ``community smells'', such as organizational silos or poor collaboration patterns, which are dysfunctional social patterns in a project's workflow \cite{tamburri2021smells, jansen2017ecosystem}. The Head of Infrastructure (P13) stated:
\begin{quote}
\textit{A common situation is that a core maintainer suddenly disappears because they changed jobs... The things they were responsible for are no longer maintained, and we have to urgently find another team to take over.}
\end{quote}

\paragraph{\texttt{Ex-Comm-3}: Collaboration Barriers} When an enterprise tries to contribute code upstream, it encounters indifference or rejection, forcing it to maintain a costly internal fork. An OSPO Governance Lead (P5) mentioned:
\begin{quote}
\textit{We once tried to `activate' a slow-responding community... But the maintainer thought our issues were `trivial' and replied, `If you want to change it, change it in your own version. I don't need to release a new version for your minor issue.'}
\end{quote}

\subsubsection{Dimension 3: Ecosystem Threats (\texttt{Ex-Eco})}
\paragraph{\texttt{Ex-Eco-1}: Regulatory Pressure} Governments enact legislation (e.g., the EU's CRA) that mandates increased software transparency and security, transforming best practices into legally binding compliance costs. An OSPO Standards Lead (P6) commented:
\begin{quote}
\textit{A few years ago, we mainly focused on license compliance. But now... The EU's CRA, the US Executive Order, both have mandatory requirements for SBOMs.}
\end{quote}

\paragraph{\texttt{Ex-Eco-2}: Competitive Disadvantage} In a domain where a project (e.g., \texttt{CUDA}) has established a \textit{de facto} standard, a newcomer will find it difficult to attract users, facing market marginalization. An engineer at an AI chip manufacturer (P2) stated:
\begin{quote}
\textit{The market leader's software ecosystem has been deeply entrenched... users don't just look at your hardware performance; they look at your software ecosystem.}
\end{quote}

\paragraph{\texttt{Ex-Eco-3}: Ecosystem Fragmentation} A successful project is ``forked'' due to commercial competition, splitting the community and diluting network effects. An OSPO Strategist (P4) analyzed:
\begin{quote}
\textit{The worst impact of recent license changes... is the fragmentation of the ecosystem. Now users have to choose between the original vendor's `source-available' version and the community-led open source fork.}
\end{quote}

\paragraph{\texttt{Ex-Eco-4}: Strategic License Changes} A commercial company unilaterally changes a project's license to a ``source-available'' license with commercial restrictions (e.g., Server Side Public License, \texttt{SSPL}). This threat is often a direct consequence of the ``Open Core'' business model, one of several archetypes identified by \cite{Duparc2022Archetypes}, where a company seeks to convert users of a free community edition into customers of a proprietary enterprise version. A license change can be a strategic lever to accelerate this conversion. This trend has been exemplified by recent high-profile relicensing events from companies like HashiCorp (Terraform) and Redis, which led to the creation of community-led, foundation-hosted forks like OpenTofu and Valkey, respectively \cite{foster2024relicensing}. This reflects a broader strategic shift from OSI-approved licenses to more restrictive ``source-available'' licenses (like \texttt{SSPL} or \texttt{BuSL}) as a commercial tactic, particularly to counter large cloud providers \cite{argentieri2024sourceavailable}. This risk attacks the legitimacy of a business model built on the original license. An OSPO Governance Lead (P5) mentioned:
\begin{quote}
\textit{When a famous data retrieval project changed its license to \texttt{SSPL}, we immediately activated our contingency plan... you must immediately evaluate alternative technology routes.}
\end{quote}

\subsection{RQ3: The Internal Vulnerability Taxonomy} 
An external threat does not exist in a vacuum; its potential for harm is determined by the specific internal weaknesses it can exploit. Our analysis revealed that the relationship between external threats and internal vulnerabilities is not a simple one-to-one mapping but a many-to-many \textit{amplifier model}. In this model, external sources act as the initial trigger, while internal vulnerabilities function as amplifiers that determine the final magnitude of the impact. The internal vulnerabilities that emerged from our grounded theory analysis find strong parallels in other empirical research. For instance, a study by \cite{kalu2025industry} on the barriers to adopting software signing identified a similar triad of technical, organizational, and human challenges, validating our multi-dimensional model of internal weakness. For example, a single external threat trigger like the \texttt{Log4Shell} vulnerability (\texttt{Ex-Tech-1}) was catastrophically amplified in organizations with poor supply chain visibility (\texttt{In-Tech-1}) and a disjointed remediation lifecycle (\texttt{In-Ops-2}). The following taxonomy, detailed in Table 5, classifies these internal amplifiers, which emerged in three layers: Strategic \& Cultural, Operational \& Process, and Technical \& Infrastructure.

\subsubsection{Dimension 1: Strategic \& Cultural Vulnerabilities (\texttt{In-Strat})}
\paragraph{\texttt{In-Strat-1}: Ambiguous Strategic Mandate} Lacking a clear, high-level mandate linked to business goals leads to disconnected efforts and wasted resources. An OSPO strategist (P4) criticized:
\begin{quote}
\textit{Many companies' open source activities are decoupled from business value... almost no one can figure out the business logic... which makes the value of open source work extremely difficult to measure.}
\end{quote}

\paragraph{\texttt{In-Strat-2}: Fragmented Governance \& Accountability} The lack of a central authority (OSPO) leads to chaotic governance and creates a ``governance vacuum'' that multiplies points of failure. An OSPO standards lead (P6) pointed out:
\begin{quote}
\textit{Without unified management standards... different teams will act inconsistently... which systematically introduces potential security, compliance, and maintenance problems.}
\end{quote}

\small
\begin{longtable}{
  >{\RaggedRight}p{0.1\textwidth} 
  >{\RaggedRight}p{0.22\textwidth} 
  >{\RaggedRight}p{0.12\textwidth} 
  >{\RaggedRight\arraybackslash}p{0.47\textwidth}
}
  
  \caption{Taxonomy of Internal Vulnerabilities}
  \label{tab:internal-vulnerabilities} \\
  
  \toprule
  \textbf{ID} & \textbf{Vulnerability Name} & \textbf{External Threats Amplified} & \textbf{Definition} \\
  \midrule
  \endfirsthead 

  \caption{Taxonomy of Internal Vulnerabilities (Continued)} \\
  \toprule
  \textbf{ID} & \textbf{Vulnerability Name} & \textbf{External Threats Amplified} & \textbf{Definition} \\
  \midrule
  \endhead 

  \bottomrule
  \multicolumn{4}{r}{\textit{Continued on next page}} \\
  \endfoot 

  \bottomrule
  \endlastfoot 

  
  \multicolumn{4}{c}{\textbf{Strategic \& Cultural (\texttt{In-Strat})}} \\
  \midrule
  \texttt{In-Strat-1} & Ambiguous Strategic Mandate & \texttt{Ex-Eco-2}, \texttt{Ex-Comm-1} & Lacks a clear, high-level strategic mandate for open source activities that is linked to business goals. \\
  \texttt{In-Strat-2} & Fragmented Governance & \texttt{Ex-Tech-4}, \texttt{Ex-Comm-3} & Governance is decentralized with no single-owner (like an OSPO), leading to inconsistent policies and a governance vacuum. \\
  \texttt{In-Strat-3} & Undefined Risk Appetite & All threats & The organization has not formally defined its risk tolerance, leading to inconsistent, \textit{ad-hoc} decision-making. \\
  \texttt{In-Strat-4} & Immature OSS Culture & \texttt{Ex-Comm-3}, \texttt{Ex-Tech-3} & Organizational culture is misaligned with secure development and ``Upstream First'' principles; developers lack ownership. \\
  \midrule
  
  \multicolumn{4}{c}{\textbf{Operational \& Process (\texttt{In-Ops})}} \\
  \midrule
  \texttt{In-Ops-1} & Immature Intake Protocol & \texttt{Ex-Tech-1}, \texttt{Ex-Tech-4} & Lacks a standardized, multi-dimensional vetting process for new components covering technical, legal, and community health. \\
  \texttt{In-Ops-2} & Disjointed Remediation Lifecycle & \texttt{Ex-Tech-1}, \texttt{Ex-Tech-2} & Lacks a closed-loop, end-to-end process for tracking identified risks from discovery through to validated remediation. \\
  \texttt{In-Ops-3} & Insufficient Compliance Mechanisms & \texttt{Ex-Tech-4}, \texttt{Ex-Eco-1} & Lacks repeatable processes for license compliance, CLA/DCO management, and systematic SBOM generation. \\
  \texttt{In-Ops-4} & Inadequate Resource Allocation & \texttt{Ex-Comm-1}, \texttt{Ex-Comm-3} & Governance functions are under-resourced and lack personnel with specialized legal, security, and community management skills. \\
  \texttt{In-Ops-5} & Lack of Feedback Loop & All threats & Lacks a systematic process to learn from incidents and failures to drive continuous improvement in governance. \\
  \midrule
  
  \multicolumn{4}{c}{\textbf{Technical \& Infrastructure (\texttt{In-Tech})}} \\
  \midrule
  \texttt{In-Tech-1} & Incomplete Supply Chain Visibility & \texttt{Ex-Tech-1}, \texttt{Ex-Eco-4} & Inability to maintain a complete, accurate, and dynamic inventory (SBOM) of all software components and their maintenance status. \\
  \texttt{In-Tech-2} & Unhardened Build Infrastructure & \texttt{Ex-Tech-2} & The development and build infrastructure (CI/CD) is not hardened against attacks, lacking features like hermetic builds or artifact signing. \\
  \texttt{In-Tech-3} & Lack of Automated Policy Enforcement & \texttt{Ex-Tech-1}, \texttt{Ex-Tech-4} & Governance relies on manual reviews rather than automated policy enforcement (security gates) within the CI/CD pipeline. \\
  \texttt{In-Tech-4} & Insufficient Intelligence Capabilities & \texttt{Ex-Comm-1}, \texttt{Ex-Eco-4} & Fails to continuously monitor the open source ecosystem for non-technical intelligence signals (e.g., project health, license changes). \\

\end{longtable}

\paragraph{\texttt{In-Strat-3}: Undefined Risk Appetite} The organization has not formally defined its risk tolerance, leading to inconsistent decision-making and ``risk management theater''. An OSPO strategist (P4) reflected:
\begin{quote}
\textit{Our internal risk assessment framework is sometimes disconnected from the real business... we might be overly concerned about license risks while ignoring more fundamental business survival issues.}
\end{quote}

\paragraph{\texttt{In-Strat-4}: Immature Open Source \& Security Culture} Organizational norms are inconsistent with secure development principles. Developers lack a sense of security ownership, and the organization fails to embrace an ``Upstream First'' mindset. An OSPO governance lead (P5) mentioned:
\begin{quote}
\textit{The business is particularly resistant to [upgrades]... if it's just from a security or compliance perspective, they find it very troublesome.}
\end{quote}

\subsubsection{Dimension 2: Operational \& Process Vulnerabilities (\texttt{In-Ops})}
\paragraph{\texttt{In-Ops-1}: Immature Intake \& Vetting Protocol} Lacking a standardized process for evaluating new components on technical, legal, security, and community health dimensions. An AI engineer (P14) described:
\begin{quote}
\textit{When the team selects dependencies, it's mainly based on personal experience... The goal is to complete tasks quickly to meet KPIs. This can introduce poorly maintained, vulnerable, or non-compliant dependencies.}
\end{quote}

\paragraph{\texttt{In-Ops-2}: Disjointed Risk-to-Remediation Lifecycle} Lacking an end-to-end process to track identified risks from discovery to remediation, creating a long ``vulnerability half-life''. This vulnerability is exacerbated by a well-documented tendency for developers to delay dependency updates, often due to a rational fear of breaking changes, a lack of awareness of available patches or known vulnerabilities, or the sheer effort required for the update \cite{kula2018outdated}. 
Even with the rise of automated tools like Dependabot, the practice of updating remains complex and fraught with challenges \cite{he2023dependabot}. This problem is magnified for large-scale library migrations \cite{he2021large, gu2023migration}, which often require specialized decision support and recommendation tools \cite{he2021migrationadvisor, he2021recommendations}. 
An OS security expert (P11) complained:
\begin{quote}
\textit{I see a project with a very low ecosystem score of 2 to 4, and I want to push for its replacement, but people will ask, `What's wrong with a score of 2 to 4?' I don't have anything convincing to persuade them.}
\end{quote}

\paragraph{\texttt{In-Ops-3}: Insufficient Legal \& Compliance Mechanisms} Lacking repeatable workflows for license scanning, CLA/DCO management, and systematic SBOM generation. An ecosystem lead (P10) pointed out:
\begin{quote}
\textit{If third-party code license requirements... are not strictly followed... it constitutes infringement and can lead to lawsuits. Therefore, we require contributors to sign a CLA.}
\end{quote}

\paragraph{\texttt{In-Ops-4}: Inadequate Resource \& Expertise Allocation} The governance function is understaffed and lacks specialized legal, security, and community management skills. An OSPO governance lead (P5) admitted:
\begin{quote}
\textit{When we evaluate non-top-tier projects, we lack rapid and effective means for in-depth assessment... This is really just throwing manpower at the problem.}
\end{quote}

\paragraph{\texttt{In-Ops-5}: Lack of a Governance Improvement Feedback Loop} The organization lacks mechanisms to systematically learn from incidents and failures, leading to repeated mistakes and a stagnant governance posture.
An ecosystem director (P10) reflected on this gap:
\begin{quote}
\textit{After a major incident like \texttt{Log4Shell}, everyone scrambles for a few weeks... but we rarely conduct a formal post-mortem to change the underlying process. We are so busy with the next fire that we don't learn from the last one, which means we often end up repeating the same mistakes.}
\end{quote}

\subsubsection{Dimension 3: Technical \& Infrastructure Vulnerabilities (\texttt{In-Tech})}
\paragraph{\texttt{In-Tech-1}: Incomplete Software Supply Chain Visibility} The inability to maintain a complete, dynamic SBOM of all software components \textbf{and their maintenance status}. Simply cataloging component versions is insufficient, as unmaintained or abandoned dependencies constitute a significant latent risk even without publicly disclosed CVEs \cite{valiev2023abandonment}. This turns every new zero-day vulnerability into a crisis. An OS security expert (P11) pointed out:
\begin{quote}
\textit{In the community, this area is quite messy; there's no such infrastructure... I feel there's some redundant investment.}
\end{quote}

\paragraph{\texttt{In-Tech-2}: Unhardened Development \& Build Infrastructure} The Continuous Integration/Continuous Delivery (CI/CD) pipelines and build systems themselves have vulnerabilities, representing the ``soft underbelly'' of an enterprise's security. A community manager (P9) mentioned:
\begin{quote}
\textit{The trust mechanism of a community's CI system can be exploited... An attacker can gain trust... and then use the elevated permissions to execute malicious code in the CI/CD process.}
\end{quote}

\paragraph{\texttt{In-Tech-3}: Lack of Automated Policy Enforcement} Relying on manual reviews instead of automated gates in the CI/CD pipeline, creating bottlenecks and fostering a culture of circumvention. An OSPO standards lead (P6) described a vulnerable state:
\begin{quote}
\textit{Although the company has established check processes, in practice they are easily bypassed... making the entire risk control system a facade.}
\end{quote}

\paragraph{\texttt{In-Tech-4}: Insufficient Intelligence \& Monitoring Capabilities} Failure to monitor the external ecosystem for non-technical signals like project health or license changes, ensuring the organization remains in a reactive, crisis-driven mode. An OSPO strategist (P4) argued:
\begin{quote}
\textit{The biggest risk is that the driving value behind a project is implicit and invisible... If you don't have expert opinions and only look at the dashboard data... what you might perceive is just a quantitative figure.}
\end{quote}

\subsection{RQ4: Risk Mitigation Framework} 
The ultimate goal of any risk governance framework is to effectively mitigate the external threats that threaten an enterprise's strategic objectives. However, many external factors—such as community dynamics, broad ecosystem shifts, or sudden market changes—are inherently uncontrollable. Direct attempts to "eliminate" these external threats are often futile.

Therefore, our framework proposes a strategic shift in perspective: from reactively firefighting external threats to proactively building an internal ``immune system''. The most effective and leveraged approach is to systematically remediate the internal vulnerabilities that act as amplifiers for these external threats. This is akin to reinforcing a house's foundation and structure; the goal is not to stop the storm, but to ensure the house can withstand it. The mitigation strategies detailed below are built upon this ``strengthening from within to defend against the without'' philosophy. Each action, while directly targeting an internal vulnerability, is ultimately aimed at enhancing resilience against specific categories of external threats.

\subsubsection{Dimension 1: Remediating Strategic \& Cultural Vulnerabilities (\texttt{In-Strat})}

To address \texttt{In-Strat-1} (Ambiguous Mandate), the OSPO must draft a charter linking all activities to the strategic objectives matrix (e.g., to justify Return on Investment, \texttt{ROI}) and secure C-level sponsorship. As P4 emphasized:
\begin{quote}
\textit{Open source activities must serve business value... Only then can we be on equal footing with product departments when requesting budget.}
\end{quote}

For \texttt{In-Strat-2} (Fragmented Governance), establishing an empowered, cross-functional OSPO is key. The establishment of an empowered OSPO is the primary institutional mechanism for mitigating fragmented governance. This is reflected in industry trends, where a significant portion of large organizations now have an OSPO to improve compliance and transparency \cite{Hendrick2024OSPO}, and in emerging organizational models designed to consolidate human infrastructure and support a coherent open source culture \cite{Ruediger2025University}. The effectiveness of a centralized OSPO in unifying policy and driving strategy is supported by numerous industry case studies from companies like Dropbox and Yahoo \cite{ospo-case-studies}, as well as theoretical frameworks for corporate open source governance \cite{decock2023corporate}. P6's experience confirms this:
\begin{quote}
\textit{Our company's governance structure is two-tiered... The top level... is responsible for decision-making; the lower level... the OSPO, is responsible for execution. This centralized model ensures consistency.}
\end{quote}
To address \texttt{In-Strat-3} (Undefined Risk Appetite), leadership must facilitate workshops to define and communicate risk tolerance, guiding trade-offs between speed and security. For \texttt{In-Strat-4} (Immature Culture), the enterprise must implement ongoing training and establish incentives, such as including open source contributions in performance reviews. As P9 mentioned,
\begin{quote}
\textit{We introduced an `Open Source Contributor' career level certification... This is more effective than any mandatory regulation.}
\end{quote}

\small
\begin{longtable}{
    >{\RaggedRight}p{0.22\textwidth} 
    >{\RaggedRight}p{0.23\textwidth} 
    >{\RaggedRight}p{0.23\textwidth}
    >{\RaggedRight\arraybackslash}p{0.23\textwidth}
}

\caption{Overview of Risk Mitigation Strategies}
\label{tab:mitigation-summary-revised} \\
\toprule
\textbf{Vulnerability} & \textbf{Core Mitigation Strategy} & \textbf{Key Initiatives \& Engineering Considerations} & \textbf{Enhanced Resilience Against External Threats} \\
\midrule
\endfirsthead

\caption[]{Overview of Risk Mitigation Strategies (Continued)} \\
\toprule
\textbf{Vulnerability} & \textbf{Core Mitigation Strategy} & \textbf{Key Initiatives \& Engineering Considerations} & \textbf{Enhanced Resilience Against External Threats} \\
\midrule
\endhead

\midrule
\multicolumn{4}{r@{}}{\textit{Continued on next page}} \\
\endfoot

\bottomrule
\endlastfoot

\multicolumn{4}{c}{\textbf{Strategic \& Cultural (\texttt{In-Strat})}} \\
\midrule
\texttt{In-Strat-1}: Ambiguous Mandate & Formulate a formal strategy linked to business goals, securing C-level sponsorship for the OSPO. & Draft OSPO charter; Link initiatives to \texttt{ROI}; Secure executive sponsor. & Improves response to \texttt{Ex-Eco-2}, \texttt{Ex-Comm-1} by ensuring long-term, strategic resource commitment. \\
\texttt{In-Strat-2}: Fragmented Governance & Establish an empowered OSPO as the single center of responsibility for open source governance. & Unify policy-making; Establish central SBOM inventory; Clarify incident response. & Enables coherent, rapid response to all ecosystem-level threats (\texttt{Ex-Eco}). \\
\texttt{In-Strat-3}: Undefined Risk Appetite & Formally define and communicate a company-wide open source risk appetite and tolerance framework. & Organize risk appetite workshops; Develop and publish decision trade-off guidelines. & Improves decision quality and speed when facing any external threat event. \\
\texttt{In-Strat-4}: Immature Culture & Implement comprehensive training and incentive mechanisms to foster a culture of ownership. & Conduct ongoing training; Include contributions in performance reviews; Promote ``Upstream First''. & Directly reduces friction with communities, mitigating \texttt{Ex-Comm-3} and improving adaptation to \texttt{Ex-Tech-3}. \\
\midrule
\multicolumn{4}{c}{\textbf{Operational \& Process (\texttt{In-Ops})}} \\
\midrule
\texttt{In-Ops-1}: Immature Intake Protocol & Establish a formal, multi-dimensional workflow for the intake and evaluation of open source components. & Design multi-dimensional evaluation model; Automate approval workflow. & Prevents intake of components with high \texttt{Ex-Tech-1}, \texttt{Ex-Tech-4}, or \texttt{Ex-Comm-1} risks. \\
\texttt{In-Ops-2}: Disjointed Remediation & Integrate risk management with development workflows, establishing automated processes for tracking and validation. & Automate ticket creation; Establish remediation \texttt{SLAs}; Implement end-to-end closed-loop tracking. & Reduces the ``half-life'' of vulnerabilities, minimizing exposure to \texttt{Ex-Tech-1} and \texttt{Ex-Tech-2}. \\
\texttt{In-Ops-3}: Insufficient Compliance & Implement automated license scanning, SBOM generation, and contributor agreement management. & Integrate scanning in CI/CD; Enforce \texttt{CLAs}/\texttt{DCOs}; Standardize SBOM generation. & Ensures compliance with legal requirements, mitigating \texttt{Ex-Tech-4} and addressing \texttt{Ex-Eco-1}. \\
\texttt{In-Ops-4}: Inadequate Resources & Staff the OSPO with dedicated, skilled personnel and budget. Use RQ1 Matrix to build the business case. & Form a cross-functional expert team; Secure a dedicated budget for the OSPO. & Provides the necessary expertise to navigate complex \texttt{Ex-Comm} and \texttt{Ex-Eco} threats. \\
\texttt{In-Ops-5}: No Feedback Loop & Institute a formal post-mortem process for all significant incidents to drive organizational learning. & Mandate blameless post-mortems; Track action items to completion; Share lessons learned. & Builds long-term resilience against all categories of recurring external threats. \\
\midrule
\multicolumn{4}{c}{\textbf{Technical \& Infrastructure (\texttt{In-Tech})}} \\
\midrule
\texttt{In-Tech-1}: Incomplete Visibility & Deploy enterprise-grade SCA tools integrated into all CI/CD pipelines to build a dynamic SBOM knowledge base. & Full-scale SCA deployment; Establish a centralized, machine-readable SBOM repository. & Enables rapid impact analysis for \texttt{Ex-Tech-1} (zero-days) and strategic response to \texttt{Ex-Eco-4}. \\
\texttt{In-Tech-2}: Unhardened Infrastructure & Implement frameworks like \texttt{SLSA} to harden CI/CD pipelines, achieving build isolation and artifact signing. & Achieve hermetic/reproducible builds; Cryptographically sign and verify all software artifacts. & Provides a strong defense-in-depth against sophisticated supply chain attacks (\texttt{Ex-Tech-2}). \\
\texttt{In-Tech-3}: No Automated Enforcement & Deploy automated gates in CI/CD pipelines, carefully balancing control with developer velocity. & ``Shift left'' security checks; Provide fast, actionable, developer-friendly feedback. & Prevents non-compliant (\texttt{Ex-Tech-4}) or vulnerable (\texttt{Ex-Tech-1}) code from entering the codebase. \\
\texttt{In-Tech-4}: Insufficient Intelligence & Establish an intelligence system combining automated tools and human analysis to monitor ecosystem dynamics. & Automate monitoring of community health metrics; Build expert networks for ``soft'' intelligence. & Provides early warning for community health decline (\texttt{Ex-Comm-1}, \texttt{Ex-Comm-2}) and ecosystem shifts (\texttt{Ex-Eco-4}). \\

\end{longtable}

\subsubsection{Dimension 2: Remediating Operational \& Process Vulnerabilities (\texttt{In-Ops})}
To mitigate \texttt{In-Ops-1} (Immature Intake Protocol), a formal, multi-dimensional, and automated workflow for vetting new components must be established. For \texttt{In-Ops-2} (Disjointed Remediation), risk management must be integrated into development workflows, automatically creating tickets from scan results and setting clear Service-Level Agreements (\texttt{SLAs}). As P11 shared,
\begin{quote}
\textit{Our biggest improvement was to directly convert SCA scan results into tickets on the development team's \texttt{Jira} board.}
\end{quote}
\texttt{In-Ops-3} (Insufficient Compliance) requires automating license scanning and SBOM generation within the CI/CD pipeline. 
To address \texttt{In-Ops-4} (Inadequate Resources), the OSPO must be staffed with a dedicated, cross-functional team and have an independent budget. Finally, for the new vulnerability \texttt{In-Ops-5} (No Feedback Loop), the enterprise must institute a formal, blameless post-mortem process for all significant incidents to drive organizational learning.

\subsubsection{Dimension 3: Remediating Technical \& Infrastructure Vulnerabilities (\texttt{In-Tech})}
To address \texttt{In-Tech-1} (Incomplete Visibility), enterprise-grade SCA tools must be deployed across all pipelines to build a dynamic, central SBOM repository. As P11 noted,
\begin{quote}
\textit{When an event like \texttt{Log4Shell} broke out, we could locate all affected assets within minutes.}
\end{quote}
For \texttt{In-Tech-2} (Unhardened Infrastructure), frameworks like Supply-chain Levels for Software Artifacts (\texttt{SLSA}) \cite{slsa_framework_2021} should be implemented to harden CI/CD pipelines with hermetic builds and artifact signing. After an attack, P9 mentioned,
\begin{quote}
\textit{We followed the \texttt{SLSA} framework to harden our build infrastructure... This greatly enhanced our ability to defend against attacks similar to the \texttt{XZ} backdoor.}
\end{quote}
To mitigate \texttt{In-Tech-3} (No Automated Enforcement), governance must be ``shifted left'' by deploying automated gates in the CI/CD pipeline, carefully balancing control with developer experience. Finally, for \texttt{In-Tech-4} (Insufficient Intelligence), an intelligence system combining automated monitoring of ``hard'' signals (e.g., community health metrics) and human analysis for ``soft'' intelligence is needed.

Beyond detection and policy enforcement, emerging technologies in Automated Vulnerability Repair (AVR) represent a more advanced mitigation frontier, directly addressing \texttt{In-Ops-2} by programmatically shortening the "vulnerability half-life." SoK studies in this domain \cite{hu2025avr, li2025sok} categorize various techniques (e.g., template-based, search-based, and learning-driven) while also highlighting current challenges, such as generating "plausible but incorrect" patches. Integrating such advanced, AI-driven repair capabilities can be seen as a forward-looking goal for organizations reaching the "Optimized" stage of governance maturity (see Section 4.4.4).

\subsubsection{Maturity-based Implementation Roadmap}
We propose a three-stage maturity implementation roadmap to guide organizations in a phased, practical approach to improving their governance capabilities.

\paragraph{Stage 1: Establish Baseline Visibility and Control}
\begin{itemize}
    \item \textbf{Core Focus:} Answering ``What do we have?'' and ``What are the rules?'' to move from an \textit{Ad-hoc} to a \textit{Managed} state.
    \item \textbf{Key Vulnerabilities to Address:} \texttt{In-Strat-1}, \texttt{In-Strat-2}, \texttt{In-Tech-1}
    \item \textbf{Specific Actions:} Formally empower the OSPO; Establish an initial component inventory (SBOM); Develop baseline usage and contribution policies.
\end{itemize}

\paragraph{Stage 2: Process Integration and Automation}
 \begin{itemize}
    \item \textbf{Core Focus:} Embedding governance into the daily workflow of developers to move from a \textit{Managed} to a \textit{Defined} state.
    \item \textbf{Key Vulnerabilities to Address:} \texttt{In-Ops-1}, \texttt{In-Tech-3}, \texttt{In-Ops-2}
    \item \textbf{Specific Actions:} Automate security and license checks in CI/CD pipelines; Establish a closed-loop remediation process for identified risks.
\end{itemize}

\paragraph{Stage 3: Proactive Hardening and Intelligence-Driven Governance}
 \begin{itemize}
    \item \textbf{Core Focus:} Shifting from passive defense to proactive risk management, moving to \textit{Measured} and \textit{Optimized} states.
    \item \textbf{Key Vulnerabilities to Address:} \texttt{In-Tech-2}, \texttt{In-Tech-4}, \texttt{In-Strat-4}, \texttt{In-Ops-5}
    \item \textbf{Specific Actions:} Implement \texttt{SLSA} practices to harden the build pipeline; Invest in open source intelligence; Deepen cultural development and establish learning feedback loops.
\end{itemize}

\subsubsection{Implementation Challenges and Considerations}
Implementing this framework is a complex engineering and organizational change effort. Practitioners will inevitably face challenges such as budget constraints, inter-departmental political resistance, and cultural inertia. We recommend leveraging the 'Strategic Objectives Matrix' (RQ1) as a tool to build the business case for governance investment. By explicitly linking each governance initiative (e.g., purchasing an SCA tool to address \texttt{In-Tech-1}) to a C-level endorsed strategic goal (e.g., `[5.1] Manage Supply Chain Risks`), OSPO leaders can more effectively secure resources and executive support.

\subsection{Validation Results: The Framework in Action}
\subsubsection{Expert Feedback and Retrospective Case Studies}
Overall, all three experts gave high praise to the framework's completeness and clarity. More importantly, the retrospective case study exercise demonstrated its powerful analytical utility. The experts were able to use the framework's language to deconstruct complex past events and uncover deeper insights.

\textbf{E1, the OSPO Lead at a FinTech giant}, used the framework to re-examine the \texttt{Log4Shell} incident \cite{spring2022log4shell}:
\begin{quote}
\textit{This framework is incredibly clarifying. At the time of \texttt{Log4Shell}, we were in a panic, seeing only the \texttt{Ex-Tech-1} (Component Vulnerability) symptom. Re-analyzing it now, the real story was our \texttt{In-Tech-1} (Incomplete Supply Chain Visibility)—it took us two weeks to even find all instances. Even deeper, it exposed our chronic \texttt{In-Ops-2} (Disjointed Remediation Lifecycle), with different teams having no consistent patching \texttt{SLA}.}
\end{quote}

\textbf{E2, the Principal Security Architect at a global cloud provider}, applied the framework to a major technology shift forced by a license change:
\begin{quote}
\textit{Your risk layering is spot on. When \texttt{Elasticsearch} changed its license, many engineers saw it as a technical problem. But your framework correctly identifies it as an \texttt{Ex-Eco-4} (Strategic License Change). It forces a higher-level discussion. The incident revealed our \texttt{In-Strat-3} (Undefined Risk Appetite); we had never formally decided how much of our core business we were willing to tie to a single commercial entity's benevolence. This framework would have enabled that crucial strategic conversation before the crisis, not during it.}
\end{quote}

\textbf{E3, the Head of R\&D at an AI chip startup}, analyzed their struggle to compete in an established ecosystem:
\begin{quote}
\textit{This is our reality on paper. We face a massive \texttt{Ex-Eco-2} (Competitive Disadvantage) against \texttt{CUDA}. Recently, we tried to contribute to a key upstream AI library to support our hardware but hit a wall of \texttt{Ex-Comm-3} (Collaboration Barriers) because key maintainers are employed by our competitor. The framework helped me articulate to my board that our bottleneck isn't just engineering; it's a community and ecosystem challenge. It exposed our \texttt{In-Ops-4} (Inadequate Resource Allocation): we don't need more compiler engineers right now, we need developer relations experts who can navigate these community dynamics.}
\end{quote}

These cases illustrate how the framework provides a structured vocabulary to diagnose problems and link
technical events to deeper strategic and operational vulnerabilities, thus confirming its analytical utility.

\section{Threats to Validity}
To ensure the rigor of our research, we identified and took measures to mitigate the following potential threats to validity.
\begin{itemize}
    \item \textbf{Construct Validity:} To ensure our constructs accurately reflect reality, we employed \textit{Investigator Triangulation}. For example, during initial coding, a disagreement arose about whether a project being commercially-led was a ``risk'' or a ``feature''. Through discussion and comparison of data from multiple interviews, we refined this concept into the specific risk of \texttt{Ex-Eco-4} (Strategic License Changes), thereby improving our framework's precision. Furthermore, our independent expert validation confirmed the framework's fidelity to real-world practice.
    \item \textbf{Internal Validity:} To ensure the truthfulness of interviewee data, we guaranteed anonymity and cross-validated core findings across participants from different roles and companies. For example, the pain point of \texttt{Ex-Eco-2} (Competitive Disadvantage) due to ecosystem lock-in, first raised by an engineer at an AI chip startup (P3), was deliberately explored and corroborated in subsequent interviews with practitioners from cloud (P7) and OS providers (P11), confirming it as a general risk.
    \item \textbf{External Validity:} We acknowledge that our sample is primarily from technology enterprises, which may limit generalizability. However, we purposefully sampled for diversity across sectors. We do not claim universal applicability, but instead discuss the framework's \textit{transferability}. We posit that technical and process vulnerabilities (e.g., \texttt{In-Tech-1}, \texttt{In-Ops-1}) are likely highly transferable to other contexts. In contrast, strategic objectives and risks related to ecosystem orchestration (e.g., goals in the ``Orchestrate'' scenario, risk \texttt{Ex-Eco-2}) are likely more context-bound to large tech firms.
    \item \textbf{Reliability:} To ensure the transparency and dependability of our research process, we have made our anonymized interview guide and the final coding structure available in our replication package, allowing other researchers to audit the analytical path from raw data to the final framework.
\end{itemize}

\section{Discussion: Dynamics, Boundaries, and Unexplored Dimensions}
\subsection{A Dynamic and Evolutionary Perspective}
While presented linearly for clarity, the framework operates as a dynamic system in practice. A key feedback loop exists where successful mitigation (RQ4) enhances organizational capabilities, which in turn enables the pursuit of more ambitious strategic objectives (RQ1). For example, a company that masters supply chain security (\texttt{In-Tech-1}) may evolve its goal from merely ``[5.1] Manage Risks'' to ``[5.5] Set Ecosystem Security Standards''. Thus, governance is not a static state but a co-evolutionary process of capability building and strategic ambition. The blurring of boundaries between some internal and external factors (e.g., is \texttt{Ex-Comm-3} an external barrier or a symptom of an internal \texttt{In-Strat-4}) is not a flaw in the model, but rather a finding that highlights the deeply intertwined nature of an enterprise and the ecosystems it inhabits.

Furthermore, a key practical utility of this model is the cross-component linkage it provides. As highlighted in our mitigation framework (Table 6), the framework components are designed to work as an integrated system. For instance, the Strategic Objectives Matrix (RQ1) is not merely a classification tool; it serves as the essential political and business tool to address the \texttt{In-Ops-4} (Inadequate Resource Allocation) vulnerability by enabling OSPO leaders to build a compelling business case (\texttt{ROI}) directly linked to C-level objectives. This interconnectedness is a key feature of the framework's practical utility.

\subsection{Boundaries and Unexplored Dimensions}
This framework provides a rational, structural model, but we must acknowledge its boundaries.
\begin{itemize}
    \item \textbf{The ``Human Element'':} All governance is ultimately enacted by people and is deeply influenced by organizational politics, departmental silos, and individual motivations. The success of a technically sound process can be determined by the leadership skills of an OSPO manager or the level of psychological safety felt by engineers. Future research could integrate theories from organizational behavior to explore these critical ``hidden variables''.
    \item \textbf{The Framework's ``Dark Side'':} We must caution against this framework being used for ``governance theater''—creating bureaucratic hurdles that stifle innovation without meaningfully reducing risk. Its implementation must be guided by a culture of enablement, not gatekeeping.
    \item \textbf{A Note on Technical vs. Organizational Vulnerabilities:} The framework deliberately focuses on vulnerabilities at the strategic, operational, and infrastructural levels, rather than cataloging specific technical flaws like individual code defects. This is a conscious choice of abstraction. In our model, a recurring pattern of code defects is not a root cause, but rather a \textit{symptom} of deeper organizational vulnerabilities. For instance, a high number of security bugs could be seen as a manifestation of an \texttt{In-Strat-4}: Immature OSS \& Security Culture or the absence of \texttt{In-Tech-3}: Lack of Automated Policy Enforcement like static analysis tools in the CI/CD pipeline. Therefore, while technical flaws are the final vector of an attack, our framework guides practitioners to address the systemic weaknesses that allow such flaws to be introduced, missed, and deployed in the first place.
    \item \textbf{A Note on Diagnostic Methodology:} This paper provides the ``what'' but not the detailed ``how'' of diagnosis. We suggest practitioners use this framework in cross-functional workshops. Developing a standardized ``Enterprise OSS Governance Maturity Assessment'' based on this framework would be a valuable practical contribution.
\end{itemize}

\section{Implications and Future Work}
\subsection{Implications for Practitioners}
This framework offers practitioners a blueprint for an integrated, continuous governance workflow. It allows them to move from diagnosis (RQ2, RQ3) to strategic planning (RQ1, RQ4) and back, creating a cycle of measurement and improvement. For developers, it clarifies the ``why'' behind governance rules. For open source maintainers, it provides insight into the risk calculus of their enterprise users.

\subsection{Implications for Researchers \& Future Work}
This study sets the agenda for several future research directions:
\begin{itemize}
    \item \textbf{From Normative to Optimized Governance:} This framework reveals inherent trade-offs (e.g., security vs. velocity). This opens an exciting new research frontier: modeling the ``Pareto Frontier'' of open source governance. Can we quantify the impact of a specific control on developer productivity? Is there an ``optimal'' level of governance beyond which returns diminish? This would move governance research from the normative (``what you should do'') to the optimized (``how much you should do'') stage.
    \item \textbf{Context-Aware Governance Theory:} Future work should not just ``validate'' this framework in other contexts (like SMBs), but explore how context systematically reshapes the framework itself. How do the resource constraints of a startup alter the hierarchy and priority of internal vulnerabilities?
    \item \textbf{The AI Revolution in Governance:} Modern AI, especially Large Language Models (LLMs), has the potential to revolutionize how this framework is implemented. We envision future work on AI agents that can: (1) perform predictive monitoring of community health by analyzing communication patterns to flag \texttt{Ex-Comm-2} risk; (2) automatically parse and analyze license changes to provide early warnings for \texttt{Ex-Eco-4}; (3) evaluate and enforce license compliance in code generated by LLMs \cite{xu2024licoeval}; (4) autonomously generate and complete missing SBOMs, thus automating the remediation of \texttt{In-Tech-1};  and (5) assist developers by generating personalized and coherent communication artifacts like release notes \cite{daneshyan2025smartnote}. Future AI agents could move beyond mere detection to automated validation, integrating techniques like those pioneered by \cite{Deng2025ChainFuzz} to generate proofs-of-concept for vulnerabilities, thus allowing governance systems to automatically prioritize genuinely exploitable threats.
    \item \textbf{Context-Aware Governance for AI/ML Supply Chains:} Our framework was derived from general enterprise software contexts, many of which are now heavily invested in AI. Future work should explicitly investigate how the unique artifacts (models, datasets) and threats (data poisoning, model inversion) of AI/ML supply chains \cite{acs2024aiml} require an \textit{extension} of our governance model. For example, empirical studies show that dependencies in LLM projects have unique vulnerability characteristics, such as higher severity and longer disclosure lags \cite{hata2024llm}. This raises the question: how must our taxonomies (\texttt{Ex-Tech}) and vulnerabilities (\texttt{In-Tech-1}) be adapted to systematically govern these new, data-driven risks?
            
    \item \textbf{The Intersection of Corporate Governance and Public Regulation:} This framework models governance as an internal, strategic best practice. However, emerging regulations like the EU's Cyber Resilience Act (CRA) \cite{leppanen2025cra} and broader debates on software liability \cite{weir2024liability} are transforming many of these ``best practices'' into legally binding obligations. A critical avenue for future research is to explore how these external regulatory pressures (\texttt{Ex-Eco-1}) reshape the priorities and trade-offs within our governance framework. For instance, does the legal threat of liability elevate the importance of \texttt{In-Ops-3} (Insufficient Compliance Mechanisms) above other strategic concerns?
\end{itemize}

\section{Conclusion}
Faced with an increasingly complex open source ecosystem, enterprises require a new mental model for risk governance. Through a rigorous grounded theory study with 15 practitioners, this paper puts forth a novel framework of enterprise open source risk governance. The core conceptual discovery is the emergent ``Objectives → Threats → Vulnerabilities → Mitigation'' logical chain (OTVM), anchored in the ``internal/external'' dualism that practitioners intuitively use to make sense of their world. Our framework formalizes this logic, providing taxonomies for external threats (\texttt{Ex-Tech}, \texttt{Ex-Comm}, \texttt{Ex-Eco}) and internal vulnerabilities (\texttt{In-Strat}, \texttt{In-Ops}, \texttt{In-Tech}).

This research provides a more holistic decision model that transcends technical checklists. Its novelty lies not in identifying unknown risks, but in synthesizing them into a coherent, analytical framework that provides a clear path from diagnosis to capability-building. It offers practitioners a toolset for strategic planning, and sets a new agenda for academic research towards optimized, context-aware, and AI-driven governance. Ultimately, this framework can help enterprises shift from a reactive ``firefighting'' mode to a proactive mode of building a resilient organizational ``immune system''.


\bibliographystyle{ACM-Reference-Format}
\bibliography{reference}

\end{document}